\documentclass{mem}
\usepackage{natbib}\usepackage{txfonts}\usepackage{balance}
\usepackage{graphicx}
\usepackage[breaklinks,dvipdfm]{hyperref}
\idline{75}{282}
\begin{document}
\def\teff{$T\rm_{eff }$}
\def\kms{$\mathrm {km s}^{-1}$}

\title{
Winter long duration stratospheric balloons from Polar regions
}

   \subtitle{}

\author{
F. \,Piacentini\inst{1}, 
A. \,Coppolecchia\inst{1},
P. \,de Bernardis\inst{1},
G. \,Di Stefano\inst{2},
A. \,Iarocci\inst{2},
L. \,Lamagna\inst{1},
S. \,Masi\inst{1},
S. \,Peterzen\inst{3},
G. \,Romeo\inst{2}
          }

\institute{
Physics Department, University of Rome La Sapienza, 
P.le Aldo Moro 5, 00185, Rome, Italy
\and
Istituto Nazionale di Geofisica e Vulcanologia, 
Via di Vigna Murata 605, 00143, Rome, Italy
\and
The ISTAR Group - Sisters Eagle Airport, Sisters, Oregon 97759 USA
\\ 
\email{francesco.piacentini@roma1.infn.it}
}

\authorrunning{Piacentini }

\titlerunning{Winter polar flights}

\abstract{
A new opportunity for astronomy, cosmology, physics, and atmospheric 
observations is the possibility to fly stratospheric payloads at 
30 - 40 km of altitude during the polar night. The absence of solar irradiation for long periods, and the extremely low temperature and stable environment of the winter stratosphere represent ideal environmental conditions while performing
astrophysical observations. 
Here we present a small and efficient platform, able to 
communicate, supply power and navigate in the harsh 
environment of the polar stratosphere.
After a balloon failure in January 2017, 
the payload was successfully flown in December 2017 from 78$^\circ$N, in 
Longyearbyen, Svalbard, Norway. 
Duration was limited to 21 hr, 
due to a southern trajectory that caused solar illumination and loss of lift. 
The instrument acquired and transmitted environmental data, and the
thermal performance of the power system are outstanding. 
The payload also included a set of attitude sensors, to monitor
payload movements.  
The information collected on this flight is essential to qualify 
the attitude control system sensors, and for the design if the thermal
and power system of the next generation  
LSPE-SWIPE telescope, devoted to the measurement of the polarization 
of the Cosmic Microwave Background radiation from a stratospheric
balloon during the Arctic polar night. 

}
\maketitle{}

\section{Introduction}
Long duration winter polar stratospheric balloon flights are 
a unique opportunity for instrumentation devoted to astronomy,
cosmology, physics, atmospheric and environmental measurements. 
For astronomical observations, winter polar ballooning offers the
possibility of a total shade from solar illumination, long duration,
due to stable temperature, a minimal residual atmospheric emission and
absorption, a wide observable sky, and a stable and
cold environment. These conditions are particularly useful for
observations in the far infrared and millimetric wavelengths, where
the signal from the Cosmic Microwave Background radiation is
observable. See e.g. \citet{2012SPIE.8452E..3FD} and \citep{2012SPIE.8446E..7AA}
\begin{figure*}[t!]
\begin{center}
{\includegraphics[width=13cm]{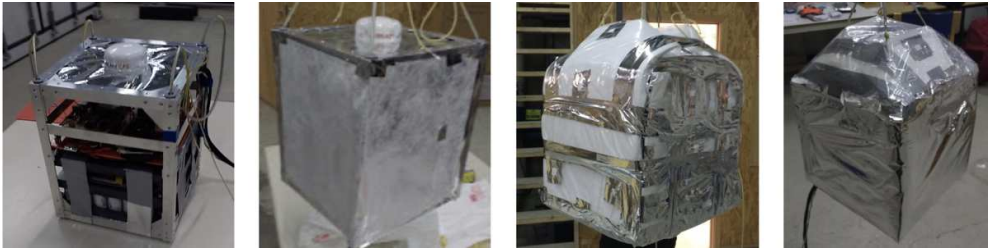}}
\end{center}
\caption{\footnotesize
Pictures of the payload. From the left: (i) assembled payload, 
without insulation; (ii) with aluminized mylar layer; (iii) with
aerogel layer; (iv) with final aluminized mylar layer. The dual
GPS+Iridium antenna is kept outside aluminized mylar layers, to
avoid electromagnetic shielding. The lower part of the payload frame
is occupied by the battery, which takes most of the volume. 
}
\label{fig:payload}
\end{figure*}

In this paper, we present a technological payload, dedicated to
testing the feasibility of winter polar long duration balloon
flights. 
The most demanding challenges of winter polar ballonning are the extreme thermal 
conditions and the power system. Temperature drops to below
$-90~^\circ$C, and a power system can't benefit from solar radiation. 
We describe how these critical aspects have been tackled, the chosen
trade-offs, and briefly describe data from the 2017 flight campaign. 

Previous experiences have been reported in~\citet{2008AdSpR..42.1633I,
  2010cosp...38.4106P, 2008MmSAI..79..940R, 2008MmSAI..79..792P, 2005ESASP.590..403P}.

\section{Instrument}
The instrument is assembled in a 30x30x30 cm cubic frame. The frame is
suspended by kevlar rope, to minimise weight and thermal
conductivity. Figure~\ref{fig:payload} illustrates the payload in its
frame. 
%
The elements of the instrument are:
\begin{itemize}
\item An Iridium modem, connected to a telemetry and termination
  electronics, based on a PIC microcontroller;
\item A power control system, to provide power regulation to the
  subsystems;
\item A thermal control system, to monitor temperatures, and activate
  heaters; 
\item An attitude sensors system, based on a Motion Processor Unit. 
\end{itemize}

The payload is completed by an independent {\em piggy back} system,
consisting in a self powered cubesat CPU and sensors  board, product of H8 robotics. 

The total weight of the system is: payload: 13.30 kg; ARGOS telemetry
system and {\em piggy back} system: 2.15~kg; flight train:
2.05~kg. Total: 17.50~kg.

\subsection{Power system}
One of the most challenging aspects of winter polar flight is the
power system. The absence of solar illumination, require a power
system fully based on batteries. 
Lithium batteries have the higher energy density (J/Kg), but
significantly loose energy at low temperature. External temperature is
expected to drop as low as $-90^\circ$C. 
Keeping batteries warm, on the other hand, requires extra power. 
It is clear the a crucial trad-off must be taken in the design of the
power system and how critical thermal insulation is.

As battery cells, we have selected the SAFT LS33600 element, combined
in a single battery of 90 cells, arranged in 15 parallel elements of 6
cells in series. A polyswitch RUE 600 limits the maximum current to
6~A. 
The assembly has a weight of 9.5 Kg, and it was designed and provided
by ELTEC Italy. At
room temperature, it provides a tension of 21.6~V, a capacity of
225~Ah, and an energy of 5508~Wh. From thermal modelling, we decided
to set the heater switch-on temperature at -40~$^\circ$C. At this
temperature, the battery provides 15.6~V, a capacity of 97.5~Ah, and
an energy of 1550~Wh, with a mere 28\% of efficiency in
energy. Nevertheless, this is the best trade-off in terms of mission
length, with a minimum intervention of the heaters.   

The system produces 1.9~W of power when heaters are off. Heaters can
provide an extra 8.8~W, with batteries at 15.6~V, for a total power of
10.7~W. With the heater switch-on temperature set at -40~$^\circ$C we
expect to have the heaters on 50\% of the time, with an estimated
duration of 12 days. If the heaters stay on all time, mission duration
will be 6 days.
\begin{figure*}[h!]
\begin{center}
{\includegraphics[width=4cm]{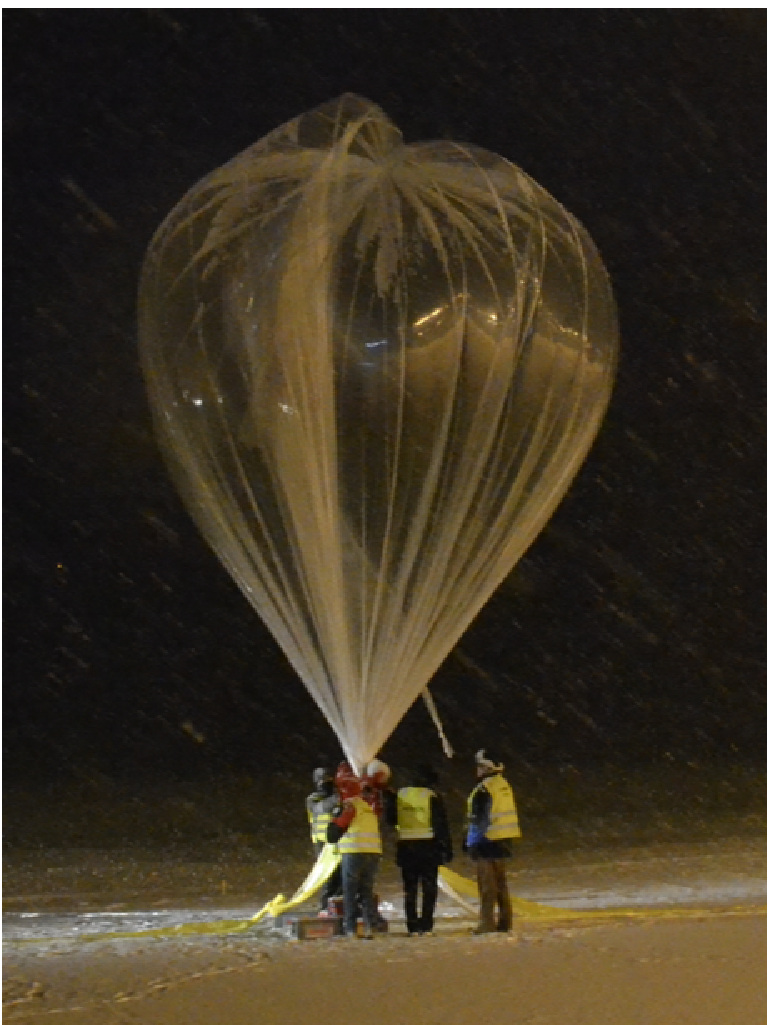}}~~{\includegraphics[width=7cm]{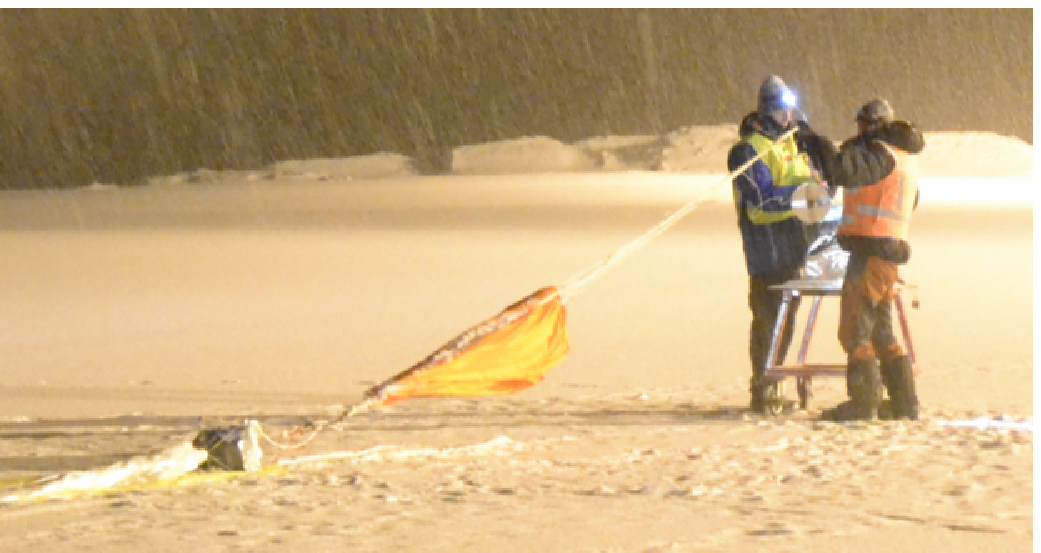}}
\end{center}
\caption{\footnotesize
Balloon and payload at launch on December 17, 2017. Soon after start
of inflation, a heavy dry snow begun. 
}
\label{fig:launch}
\end{figure*}

\subsection{Thermal insulation system}
External temperature is expected to drop to $-90~^\circ$C
during flight, while batteries must be kept at  $-40~^\circ$C. 
Thermal insulation is based on an Aerogel layer. The selected model is
Cryogel x201, which is expected to have a thermal conductivity of
8~mW/m/K in the temperature range between $-90$ and $-40~^\circ$C, at a
pressure of 10~mbar. The adopted aerogel consists of a 30x30~cm panels,
with a thickness of 25~mm. 

In order to further reduce the thermal conductivity 6 layers of
aluminized maylar have been added, 3 on the internal side and 3 on the
external side of the aerogel. 

\subsection{Telemetry system}
The telemetry system is based on an Iridium Short Bust Data (SBD),
model SBD 9202N, with a dual GPS/Iridium active antenna. 
Short messages can be sent from payload to ground, and from ground to
payload. 
The payload sends a message every 120 seconds. The message contains
coordinates and time from GPS, and data from all the sensors
onboard. More frequent data are saved on a solid state memory card
on-board. Data are received as e-mail messages on ground and converted
by the ground segment into data-stream. Each e-mail message also
contains coordinates as obtained from triangulation of Iridium
receiving satellites. This provided redundancy on the payload
tracking. 

An other redundant tracking system is connected to the balloon, and
consists of 
an ARGOS-System tracker. This tracker is only activated after
termination by a pull-and-activate system. In this way, a heavy
battery system is not required for the ARGOS tracker, and 8 battery
cells (2 series of 4 cells in parallel) are enough to provide power
after termination, and locate the balloon for at least 24 hrs. 
\begin{figure*}[t!]
\begin{center}
{\includegraphics[width=7cm]{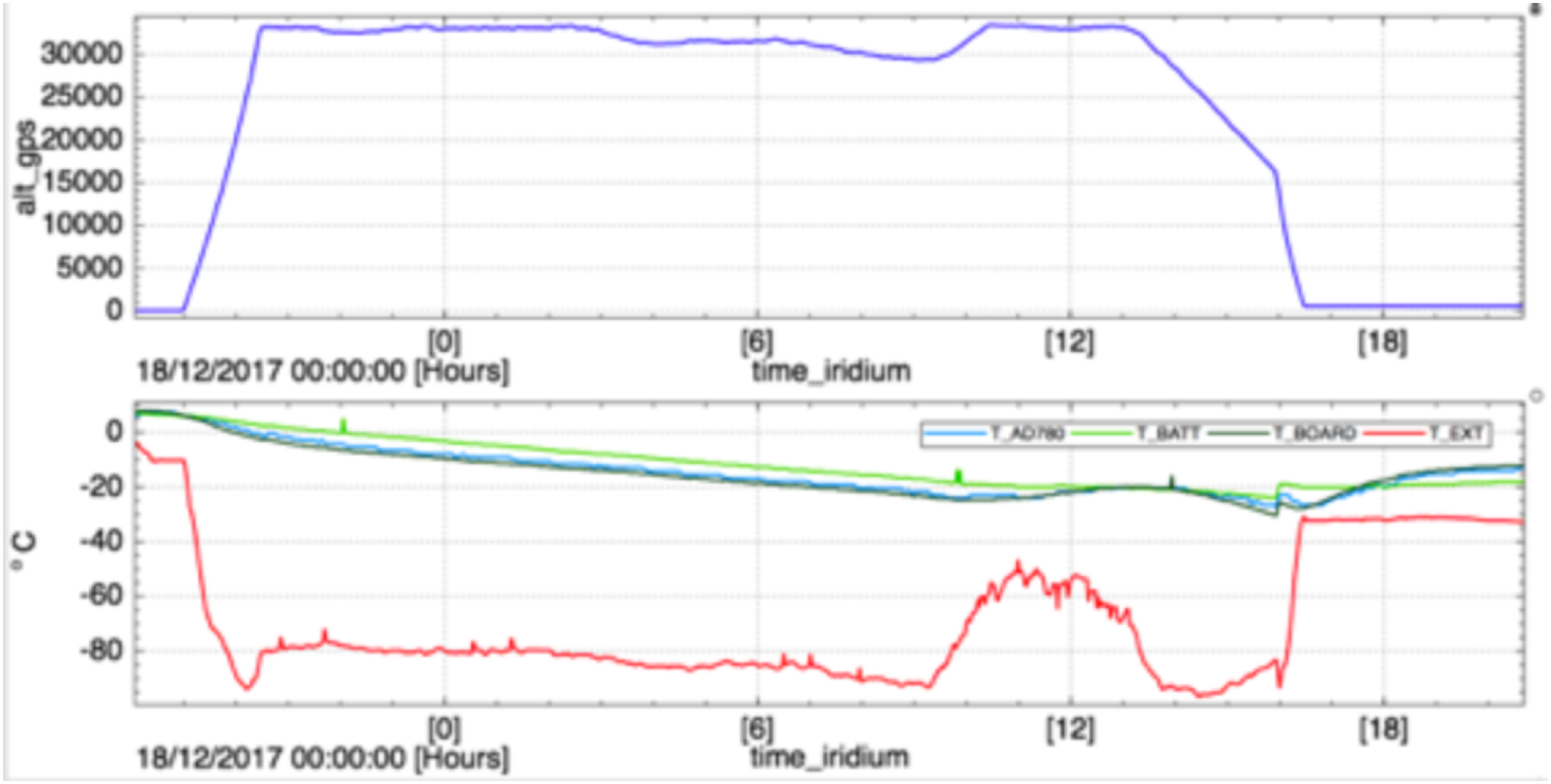}}~~{\includegraphics[width=6cm]{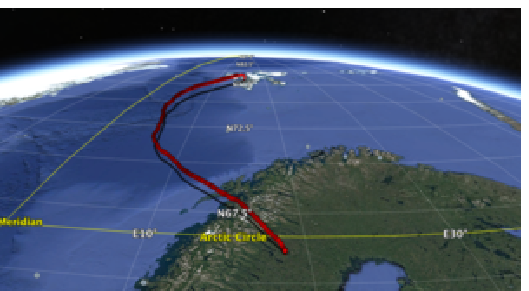}}
\end{center}
\caption{\footnotesize
Left: Altitude and temperature data vs UTC time. 
In red the outside temperature, in light green the battery
temperature, in dark green the electronics board temperature, and in
light blue the power control unit temperature. Right: 3D trajectory of the payload during 21 hrs of flight. 
}
\label{fig:data}
%
%
\end{figure*}
\subsection{Sensors}
The payload is equipped with environmental and attitude sensors. 

Environmental sensors include:
GPS (longitude, latitude, altitude); 
battery tension sensor; 
heater activation sensor; 
temperature sensors:
a PT100 for external temperature, 
an AD780 for power control unit temperature, 
an AD590 for electronics board temperature, 
an other AD590 on battery pack; 

The payload is also equipped with a Flight Controller Board (CRIUS All
In One PRO unit) based on Arduino platform. The sensors present on the
board are: 6-axis gyro/accel with Motion Processor Unit, 3-axis
digital magnetometer, high precision altimeter with metal cup and
thermometer. 
The system logs data on an external SD memory,
ans sends a fraction of the data to the Iridium SBD board by
serial communication link.

\section{Flight and performance}
For Arctic winter flights Longyearbyen (Svalbard, Norway), at
78$^\circ$N, represents an ideal location to guarantee total darkness
during operation.  
The payload was first flown in January 2017, but the balloon suffered
a failure during launch. The payload never lifted, and it was
recovered undamaged. A second successful launch was organized in
December 2017. Both launches and campaign management were orchestrated
by The ISTAR Group\footnote{S. Peterzen}. 

\subsection{Balloon}
The balloon is an Aerostar International, model SF4-0.327-.6/0-TA,
with a volume of 9288 cubic meters. It weights 37~kg and
can lift up to 20~kg. Figure~\ref{fig:launch} shows pictures of
balloon and payload during launch. An heavy dry snow didn't disturbe
operations, 

\subsection{Flight}
Balloon was inflated in the Longyearbyen airlift, after coordination
with local Air Trafic Control. External temperature was $-11~^\circ$C.
Inflation took about 30 minutes. During inflation a heavy dry snow hit
the airport, without damage to operations.  
The Balloon lifted at 18:58 UTC, 19:58 local time, on December 17,
2017. 
Ascent took 93 minutes, up to 33200 meters, with an ascent rate of 5.9
m/s. The ascent stopped at 33200 meters, without overshooting. 

During ascent, temperature dropped to $-94~^\circ$C, and then stabilised at
$-80~^\circ$C. 
At 2 AM UT (Dec 18) temperature and altitude started dropping again. 
Stratospheric wind pushed the balloon toward SW. At about 73.5N, 6.0E
the balloon turned towards SE. 
At 9AM the Sun hit the balloon causing a rise in temperature and
altitude. 
External temperature raised to above $-60~^\circ$C.  The maximum altitude was
33480, with a sharp stop probably due to balloon venting.  
Temperature and altitude remained stable until 13 UT. 
During cruise, the internal temperature never dropped below $-32~^\circ$C, and
the heater, set at $-40~^\circ$C, never switched on.  

At 13 UT, on December 18, 2017, solar illumination ceased, the
temperature dropped again with a minimum of $-96\;^\circ$C. 
The altitude dropped too, with a constant descent at a speed of 1.7 m/s. 
At 15:56UT, after 21 hours of flight the altitude was ad 16480
meters, with constant descent rate. We 
decided to terminate the flight, after contact with the Swedish Air Traffic Control. 
Termination worked perfectly, and the payload dropped smoothly in an
inhabited  area of Lappland, in the region of Arvidsjaur, just a few
km North of  Jerfojaur, not far from a country road. 

Altitude, temperature data and trajectory are reported in figure~\ref{fig:data}.

\section{Conclusions}
Long duration winter polar flights are becoming a real possibility for
instrumentation requiring a dark environment for long time. 

The payload, launched on December 2017 from Longyearbyen (Norway) 
suffered stratospheric wind towards South, which caused early
termination. Nevertheless all the subsystems worked without failures in the
harsh condition. In particular the thermal insulation confirmed
performance declared by the producer, and granted an efficient battery
power system. This opens the path to a larger
power system, such as the one needed for the forthcoming LSPE-SWIPE
balloon experiment, planned for 2020~\citep{2012SPIE.8452E..3FD}.

\begin{acknowledgements}
The experiment has been granted mainly by CNR-PNRA (Programma
Nazionale di Ricerca in Antartide, grant 14-00027), with support from Sapienza -
University of Rome, and INGV (Istituto Nazionale di Geofisica e
Vulcanologia). 
|
We acknowledge the support of the Italian Space Agency, LNS-Spitzbergen. 
Special thanks to Get-Lost for recovering the full equipment. 
\end{acknowledgements}

\bibliographystyle{aa}

\end{document}